# Thermal Conductivity Estimation of Thermoelectric Materials with Uncertainty Quantification Using Bayesian Physics-Informed Neural Networks

Hyeonbin Moon[1†], Hanbin Cho [1†], Wabi Demeke[1], Byungki Ryu[2*] and Seunghwa Ryu[1,3*]

**Affiliations**

[1] Department of Mechanical Engineering, Korea Advanced Institute of Science and Technology (KAIST), 291 Daehak-ro, Yuseong-gu, Daejeon 34141, Republic of Korea

[2] Energy Conversion Research Center, Korea Electrotechnology Research Institute (KERI), 12 Jeongiui-gil, Seongsan-gu, Changwon-si, Gyoengsangnam-do 51543, Republic of Korea

[3] KAIST InnoCORE PRISM-AI Center, Korea Advanced Institute of Science and Technology (KAIST), Daejeon, 34141, Republic of Korea

[†]These authors contributed equally to this work

[*]Corresponding author: ryush@kaist.ac.kr (Seunghwa Ryu), byungkiryu@keri.re.kr (Byungki Ryu)

**Abstract**

Characterizing the temperature-dependent thermal conductivity is challenging because the property varies strongly with temperature and reliable heat flow measurement, not just temperature sensing, is difficult under experimental conditions. Here, we present a physics-informed deep learning framework that infers conductivity solely from sparse electric potential measurements. We first develop a deterministic physics-informed neural network (PINN) that embeds coupled thermoelectric transport equations as soft constraints, enabling simultaneous recovery of spatial temperature, voltage, and conductivity profiles without temperature data. The deterministic PINN achieves accurate inference under noise-free conditions, yet its predictions degrade when measurement noise is introduced. To address this, we extend the framework to a Bayesian PINN, which models network parameters probabilistically and employs Hamiltonian Monte Carlo (HMC) sampling for posterior inference. This extension produces robust thermal conductivity estimates and, importantly, provides credible intervals that quantify uncertainty from sparse and noisy data. Numerical experiments confirm that the Bayesian PINN not only preserves predictive accuracy under noise but also reveals inference bias and enables uncertainty-aware interpretation of material properties. Together, the deterministic and Bayesian formulations establish a scalable and generalizable alternative to conventional methods for determining temperature-dependent properties, offering physics-consistent and risk-aware property inference for thermoelectric systems and other functional materials where direct temperature sensing is impractical.

1. **Introduction**

Thermal materials can be broadly classified into two categories: thermal management materials, which dissipate or move heat efficiently (e.g., for refrigeration and cooling), and thermoelectric (TE) materials, which convert heat into electricity. Among these, TE materials have attracted significant attention for their ability to directly convert heat into electricity and vice versa, enabling applications in waste heat recovery and power generation in environments where conventional energy conversion is difficult [1-5]. The efficiency of thermoelectric devices is quantified by the dimensionless figure of merit $zT$, defined as $zT = \frac{\alpha^2 \sigma}{k}T$, where $\sigma$ is the electric conductivity, $\alpha$ is the Seebeck coefficient and $k$ is the thermal conductivity; all of which are highly temperature dependent [6-7]. Here, $T$ denotes the absolute temperature. Accurate estimation of these temperature-dependent properties is essential for optimizing device performance [8-9], guiding materials selection [10], and calibrating computational models [11].

While $\sigma(T)$ and $\alpha(T)$ can be reliably measured with established techniques such as a steady-state differential method or four-probe setups [12-13], determining $k(T)$ remains far more challenging. In one dimension, thermal conductivity is obtained from $k = Q/\Delta T$, where $\Delta T$ can be measured accurately, but the net heat flow $Q$ is difficult to quantify due to parasitic losses (e.g., contact resistance, radiation) [14], environmental interference and microstructural heterogeneity [15]. In addition, maintaining stable boundary temperatures $T_1$ and $T_2$ requires significant time to mitigate fluctuations. Conventional practice determines $k(T)$ using standardized thermophysical methods such as the heat flow meter for steady-state measurements, laser flash analysis (LFA) where diffusivity is combined with heat capacity and density, and the modified transient plane source TCi for small or single-sided samples [16]. In addition, $k(T)$ can be reconstructed by solving an inverse heat-conduction problem from transient temperature histories using the network simulation method [17]. Complementing these, widely used transient methods such as LFA [18-19] and time-domain thermoreflectance (TDTR) [14, 20], estimate $k$ at each temperature setpoint under a locally constant-$k$ assumption within the fitting window; the temperature dependence is then obtained stepwise by repeating the measurement across setpoints. As will be discussed in a following section, this constant-$k$ assumption poses risks for robustness because it neglects the $\nabla k \cdot \nabla T$ contribution inherent in the divergence form $\nabla \cdot (k(T)\nabla T)$. However, in practice, internal gradients, boundary uncertainties, and interface resistance often make these estimates unreliable; [21-23] moreover, attaching measurement electrodes can locally withdraw heat and depress the nearby temperature, further biasing the measurements [24]. While some studies have attempted to circumvent these limitations using indirect techniques or model-based fitting, they generally require extensive calibration and are sensitive to measurement noise and boundary condition uncertainty [25-26].

In parallel, recent advances in data-driven modeling, including machine learning (ML) and physics-informed neural networks (PINN), have shown potential in approximating material behavior from indirect or incomplete data [27-31]. However, purely data-driven models tend to lack physical consistency and are prone to overfitting in sparse or noisy measurement regimes [32-34]. To address these limitations, probabilistic approaches leveraging Bayesian inference for uncertainty quantification are utilized to produce calibrated predictive distributions and improve robustness under sparse or noisy data. Representative examples include Bayesian neural networks (BNN) for modeling and interpreting material behavior [35-36], Bayesian PINN for flow field inference [37] as well as for tackling inverse problems [38-40]. Motivated by these developments, we adopt a probabilistic approach for the inverse recovery of $k(T)$ in TE systems using only sparse electric potential measurements.

In this study, we present a physics-informed deep learning framework for estimating the temperature-dependent thermal conductivity $k(T)$ of TE materials without requiring direct temperature measurements. Assuming that the $\sigma(T)$ and $\alpha(T)$ are known and noise-free, we formulate an inverse problem that infers $k(T)$ using only sparse electric potential measurements collected under multiple current loading conditions. We first apply a standard PINN to this setting and find that it accurately infers $k(T)$ when the measurement data is noise-free. However, its performance degrades significantly under measurement noise, showing sensitivity to data quality and overfitting even with careful tuning of the loss balance. To address this limitation, a Bayesian PINN framework is adopted with pre-training, and uncertainty quantification is incorporated by modeling the network parameters probabilistically. While this approach does not eliminate the inherent ill-posedness of the inverse problem, it significantly improves the robustness of the inference under noisy conditions and enables estimation of credible intervals that quantify uncertainty of $k(T)$. This allows not only more stable predictions but also clearer insight into how confidently $k(T)$ can be inferred. Through this analysis, we demonstrate that Bayesian PINN offers a practical and interpretable solution for estimating thermal conductivity under realistic experimental constraints.

The remainder of this paper is organized as follows: **Section 2** presents the governing equations and problem formulation, **Section 3** describes the PINN and Bayesian PINN frameworks, **Section 4** discusses the results, and **Section 5** concludes the work.

## 2. Problem formulation

This section presents the steady-state governing equations for thermoelectric materials and outlines the inverse problem for estimating the temperature-dependent thermal conductivity $k(T)$.

### 2.1) Governing equations

The coupled behavior of charge and heat transport in thermoelectric materials is governed by the following steady-state conservation equations:

$$\nabla \cdot \boldsymbol{J} = 0 \tag{1a}$$

$$\nabla \cdot \boldsymbol{q} + \nabla V \cdot \boldsymbol{J} = 0 \tag{1b}$$

where $\boldsymbol{J}$, $\boldsymbol{q}$, and $V$ denote the electrical current density, heat flux, and electric potential respectively. These quantities are related through the constitutive relations:

$$\boldsymbol{J} = -\sigma(T)(\nabla V + \alpha(T)\nabla T) \tag{1c}$$

$$\boldsymbol{q} = -k(T)\nabla T + \alpha(T)T\boldsymbol{J} \tag{1d}$$

These equations capture the combined effects of charge conduction, Seebeck, Peltier, and Joule heating. Due to the strong temperature dependence of all three material properties, the governing equations are inherently nonlinear and strongly coupled.

To simplify the analysis, we adopt a one-dimensional (1D) approximation [41], which can be physically justified in typical pi- shape bulk thermoelectric devices where both heat and charge flow predominantly along the axial direction. Under this assumption, the governing equations reduce to:

$$\frac{dJ}{dx} = 0 \tag{2a}$$

$$\frac{dq}{dx} + \frac{dV}{dx}J = 0 \qquad (2b)$$

$$J = -\sigma(T)\left(\frac{dV}{dx} + \alpha(T)\frac{dT}{dx}\right) \qquad (2c)$$

$$q = -k(T)\frac{dT}{dx} + \alpha(T)TJ \qquad (2d)$$

These equations form the basis of the inverse problem considered in this work, characterizing the coupling between electric potential, temperature, and the underlying material properties.

**2.2) Inverse problem**

The objective of this study is to estimate the unknown temperature-dependent thermal conductivity $k(T)$, along with the temperature and electric potential, $T(x)$ and $V(x)$, using a limited set of electric potential measurements. The problem is defined on a one-dimensional thermoelectric domain of length $L = 100mm$, with the spatial coordinate $x \in [0, L]$, and is subject to the following boundary conditions:

$$T(x = 0) = T_h, \qquad T(x = L) = T_c, \qquad V(x = 0) = 0 \qquad (3a)$$

$$-\sigma(T)\left(\frac{dV}{dx} + \alpha(T)\frac{dT}{dx}\right)\bigg|_{x=L} = J_n \qquad (3b)$$

Here, $T_h = 650K$ and $T_c = 350K$ denote the temperatures at the hot and cold ends, respectively. Eq. (3a) imposes Dirichlet boundary conditions on both temperature and electric potential, while Eq. (3b) defines a Neumann boundary condition corresponding to an externally applied current density at the cold side.

To enhance the identifiability of the inverse problem, we consider four distinct current loading cases, corresponding to externally applied current densities of 5.0, 10.0, 25.0, 50.0

$\mu A/mm^2$, denoted as $J_n$ for n = 1,2,3,4 For each current case, electric potential measurements are taken at three spatial locations within the domain, specifically at $x_m = \{L/3, 2L/3, L\}$ for $m = 1,2,3$. This configuration yields a total of 12 electric potential observations across all loading conditions. The use of multiple loading scenarios induces diverse system responses, thereby enhancing the constraints on the inverse problem and improving the recoverability of the unknown function $k(T)$. Unlike scalar parameter estimation, the present formulation seeks to infer a continuous function, substantially increasing the dimensionality of the solution space and rendering the problem ill-posed under sparse data conditions.

To estimate $k(T)$, the inverse problem is formulated as the following constrained optimization:

$$\min_{T(x),V(x),k(T)} \sum_{n=1}^{4}\sum_{m=1}^{3} |h^{(n,m)}|^2 \tag{4a}$$

subject to:

$$f^{(n)}(x) = 0, \quad \forall x \in [0,L], \quad n = 1,2,3,4 \tag{4b}$$

$$g^{(n)}(x) = 0, \quad \forall x \in [0,L], \quad n = 1,2,3,4 \tag{4c}$$

The residual terms $f^{(n)}(x)$, $g^{(n)}(x)$ and $h^{(n,m)}$ in Eqs. (4a)–(4c) are defined as follows:

- Current conservation residual

$$f^{(n)}(x) = J(x;J_n) - J_n \tag{4d}$$

- Energy conservation residual

$$g^{(n)}(x) = \frac{dq(x;J_n)}{dx} + \frac{dV(x;J_n)}{dx} J(x;J_n) \tag{4e}$$

- Measurement residual

$$h^{(n,m)} = V(x_m; J_n) - V_{m\,eas}(x_m; J_n) \tag{4f}$$

Here, $J(x; J_n)$ and $q(x; J_n)$ denote the current density and heat flux evaluated under the $n$-th applied current condition. The indices $n$ and $m$ correspond to the current loading case and the spatial measurement location, respectively. The current conservation residual $f^{(n)}(x)$ is derived from Eq. (2a) and the boundary condition in Eq. (3b), while the energy conservation residual $g^{(n)}(x)$ corresponds to Eq. (2b). The measurement residual $h^{(n,m)}$ represents the difference between predicted and observed electric potentials at the measurement points.

Because measurement noise is unavoidable in practical experiments, we also consider the case in which the 12 voltage measurements are corrupted by independent Gaussian noise:

$$V_{m\,eas}(x_m; J_n) = \tilde{V}_{m\,eas}(x_m; J_n) + \varepsilon \quad \text{where} \quad \varepsilon \sim N(0, \sigma_\varepsilon^2) \tag{5}$$

Here, $\tilde{V}_{m\,eas}$ is the noise-free ground-truth value, and $\sigma_\varepsilon$ is the standard deviation of the noise. In this study, $\tilde{V}_{m\,eas}$ is derived from high-fidelity simulations, discussed in **Section 4**.

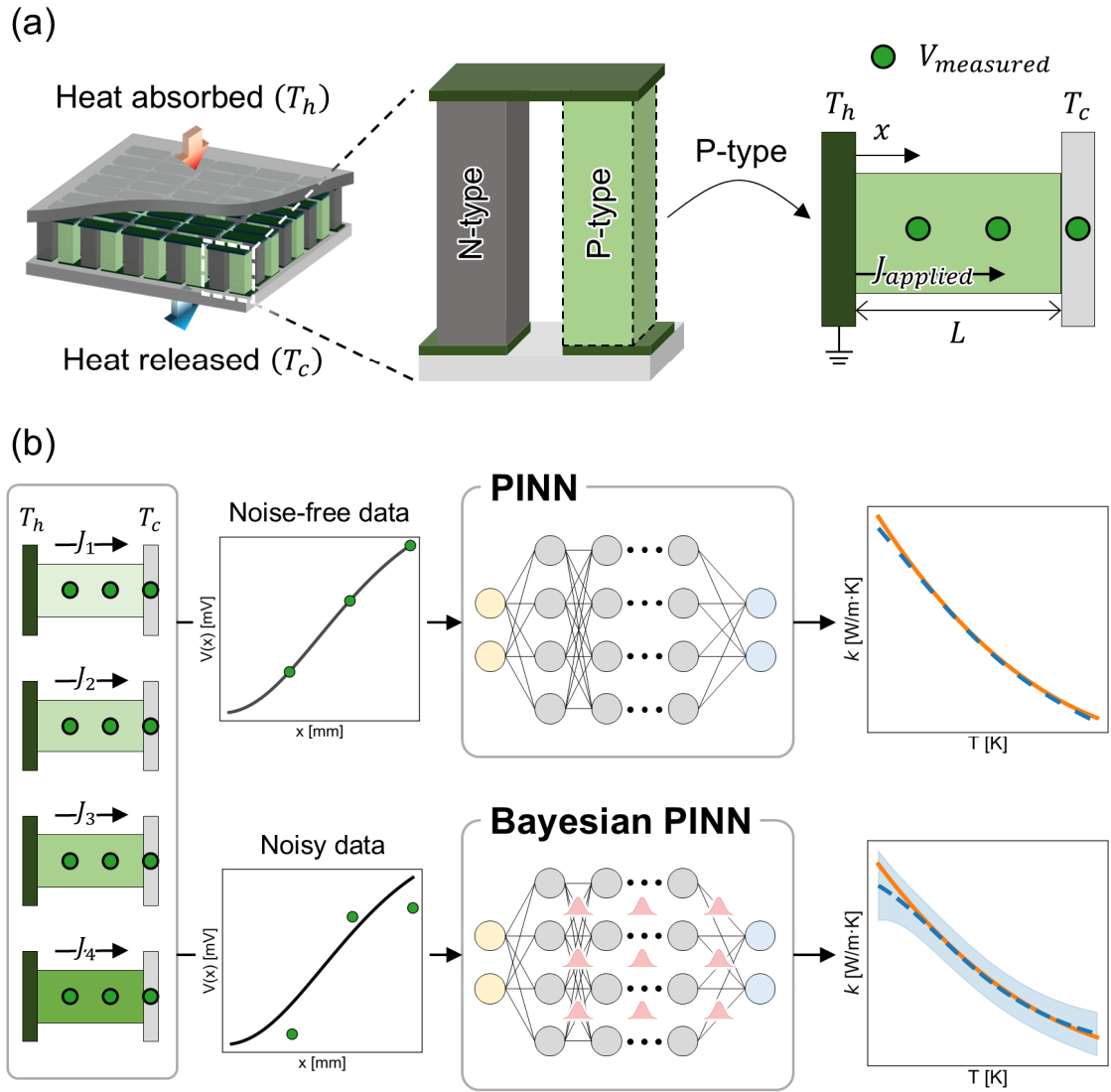

**Fig. 1. Schematic overview of the problem setup and the proposed inverse framework.** (a) Architecture of the thermoelectric device and the reduction of the P-type leg to a 1-D modeling domain with isothermal boundaries. A constant current density flows from hot to cold side, and three voltage probes are placed along the leg. (b) For four different loading conditions ($J_1 - J_4$), voltage values were measured. Noise-free data are used with a PINN to perform deterministic inference of $k(T)$. For data containing noise, a Bayesian PINN performs probabilistic inference of $k(T)$, providing a predictive mean with credible intervals that quantify the uncertainty.

## 3. Methodology

This section presents the formulations of the PINN and the Bayesian PINN, which are used to solve the inverse problem of estimating the temperature-dependent thermal conductivity from sparse electric potential measurements, as outlined in **Section 2.2**.

### 3.1) Physics-Informed Neural Networks

We employ a neural network composed of three fully connected multilayer perceptrons (MLPs) to infer the spatial profiles of $T(x)$, $V(x)$, and $k(T)$. The neural networks for $T(x)$ and $V(x)$, denoted as $NN_T(x; J_n)$ and $NN_V(x; J_n)$, take the spatial coordinate $x$ and applied current density $J_n$ as inputs. The thermal conductivity network $NN_k(T)$ takes the predicted temperature as input to model the temperature dependence.

To satisfy physical boundary conditions, the outputs are embedded into the following hard-constraint formulations:

$$T(x; J_n) = NN_T(x; J_n)x(L - x) + \frac{T_c - T_h}{L}x + T_h \tag{6a}$$

$$V(x; J_n) = NN_V(x; J_n)x \tag{6b}$$

$$k(T) = \ln(1 + \exp(NN_k(T))) \tag{6c}$$

Eqs. (6a) and (6b) impose Dirichlet boundary conditions on temperature and electric potential as defined in Eq. (3a). The Softplus function in Eq. (6c) ensures that $k(T)$ remains strictly positive.

The PINN is trained by minimizing a loss function that combines physics and data terms:

$$\text{Loss} = \lambda \frac{1}{N} \sum_{n=1}^{4} \left( \sum_{i=1}^{N} |f^{(n)}(x^{(i)})|^2 + \sum_{i=1}^{N} |g^{(n)}(x^{(i)})|^2 \right) + \sum_{n=1}^{4} \sum_{m=1}^{3} |h^{(n,m)}|^2 \quad (7)$$

Here, $x^{(i)}$ denotes the $i$-th collocation point, and $N$ is the total number of collocation points. The first two terms correspond to the physics loss, enforcing Eqs. (2a) and (2b) through the residuals $f^{(n)}$ and $g^{(n)}$, respectively. These are computed by differentiating the inferred quantities $T(x)$, $V(x)$ and $k(T)$. The final term corresponds to the data loss, which penalizes the discrepancy between the predicted and measured electric potentials through $h^{(n,m)}$. The weighting parameter $\lambda$ balances the relative influence of the physics loss and data loss during training.

To enhance numerical stability and facilitate effective training of the PINN model, all governing equations and material parameters are expressed in nondimensional form using appropriate reference quantities. Specifically, reference values for length $L_{ref}$, thermal conductivity $k_{ref}$, electrical conductivity $\sigma_{ref}$ and temperature $T_{ref}$ are used to normalize the physical variables. Derived reference quantities include the Seebeck coefficient $\alpha_{ref} = \sqrt{k_{ref}/(\sigma_{ref} T_{ref})}$, electric potential $V_{ref} = \alpha_{ref} T_{ref}$ and current density $J_{ref} = \sigma_{ref} V_{ref} / L_{ref}$. The nondimensional variables are defined as:

$$x^* = \frac{x}{L_{ref}}, \quad T^* = \frac{T}{T_{ref}}, \quad V^* = \frac{V}{V_{ref}}, \quad k^* = \frac{k}{k_{ref}},$$
$$\sigma^* = \frac{\sigma}{\sigma_{ref}}, \quad \alpha^* = \frac{\alpha}{\alpha_{ref}}, \quad J^* = \frac{J}{J_{ref}} \quad (8)$$

All physical quantities are nondimensionalized prior to training. After inference, the predicted results are converted back to their dimensional forms by multiplying the corresponding reference values.

## 3.2) Bayesian Physics-Informed Neural Networks

While PINN enforce governing physical laws during training and yield accurate point estimates, they do not account for uncertainties arising from limited, noisy, or sparse data, nor from inherent ambiguities in model parameters. Consequently, such deterministic models may produce overconfident or unreliable predictions, especially in weakly informed regimes where the available observations provide limited information. To address these limitations, a Bayesian PINN framework is adopted, in which the trainable parameters $\boldsymbol{\theta}$ of the neural networks are treated as random variables. This approach enables the estimation of the temperature-dependent thermal conductivity $k(T)$ along with associated credible intervals, thereby providing robust and interpretable predictions under uncertainty.

Bayesian PINN training is conducted in two stages. The first stage is deterministic pre-training, performed prior to probabilistic inference to obtain a well-localized initialization and improve convergence [34]. Then, inference is performed on the posterior distribution $p(\boldsymbol{\theta}|D)$, where $D$ denotes the observed data. According to Bayes' theorem, the posterior is proportional to the product of the likelihood and the prior:

$$p(\boldsymbol{\theta}|D) = \frac{p(D|\boldsymbol{\theta})p(\boldsymbol{\theta})}{p(D)} \qquad (9)$$

where $p(\boldsymbol{\theta}|D)$ denotes the posterior distribution, $p(D|\boldsymbol{\theta})$ is the likelihood function, reflecting the data fit under parameters $\boldsymbol{\theta}$. $p(\boldsymbol{\theta})$ represents the prior distribution over the model parameters. $p(D)$ is the marginal likelihood or evidence, which serves as a normalizing constant but is not needed during sampling.

Since the posterior distribution is typically high-dimensional and intractable, direct sampling is not feasible [42-43]. To efficiently explore the posterior, we employ the NUTS [44], a variant of HMC that automatically adapts the trajectory length by building balanced binary trees and applying a stopping criterion based on a "no U-turn" condition. NUTS augments the parameter space with an auxiliary momentum variable $\boldsymbol{p}$, leading to a joint density defined as:

$$p(\boldsymbol{\theta}, \boldsymbol{p}) \propto \exp(-H(\boldsymbol{\theta}, \boldsymbol{p})) \tag{10}$$

where $H(\boldsymbol{\theta}, \boldsymbol{p})$ is the Hamiltonian function, defined as the sum of the potential and kinetic energy:

$$H(\boldsymbol{\theta}, \boldsymbol{p}) = U(\boldsymbol{\theta}) + K(\boldsymbol{p}) \tag{11}$$

The potential energy $U(\boldsymbol{\theta})$ is defined in terms of the negative log-posterior:

$$U(\boldsymbol{\theta}) = -\log(p(D|\boldsymbol{\theta})p(\boldsymbol{\theta})) = -\log(p(D|\boldsymbol{\theta})) - \log p(\boldsymbol{\theta}) \tag{12}$$

The kinetic energy $K(\boldsymbol{p})$ is commonly defined as a quadratic form:

$$K(\boldsymbol{p}) = \frac{1}{2}\boldsymbol{p}^T M^{-1} \boldsymbol{p} \tag{13}$$

with a symmetric positive-definite mass matrix $M$ that is initialized to the identity and adapted during warmup. Hamiltonian dynamics are numerically integrated with the symplectic leapfrog algorithm that preserves volume and approximately conserves the Hamiltonian. Instead of a fixed trajectory length, NUTS recursively builds a balanced binary tree of leapfrog steps forward and backward, stopping automatically when the "no U-turn" condition is met or a maximum tree depth is reached. Specifically, the tree expansion terminates when

$$(\boldsymbol{\theta}^+ - \boldsymbol{\theta}^-) \cdot \boldsymbol{p}^- < 0 \text{ or } (\boldsymbol{\theta}^+ - \boldsymbol{\theta}^-) \cdot \boldsymbol{p}^+ < 0 \tag{14}$$

where $\boldsymbol{\theta}^-$ and $\boldsymbol{\theta}^+$ denote the leftmost and rightmost positions in the current trajectory, and $\boldsymbol{p}^-$ and $\boldsymbol{p}^+$ are the corresponding momenta. The next state is then sampled uniformly from the set of valid states explored during this trajectory, ensuring detailed balance. After a series of leapfrog steps, a proposal $(\boldsymbol{\theta}', \boldsymbol{p}')$ is generated. The acceptance statistic is defined as

$$\alpha = \min(1, \exp[-H(\boldsymbol{\theta}', \boldsymbol{p}') + H(\boldsymbol{\theta}, \boldsymbol{p})]) \qquad (15)$$

A slice variable $u \sim \mathcal{U}(0, \exp[-H(\boldsymbol{\theta}, \boldsymbol{p})])$ is drawn at the beginning of each trajectory. The next state is selected uniformly from the subset of visited states whose joint density satisfies $\exp[-H(\boldsymbol{\theta}', \boldsymbol{p}')] \geq u$ and that meet the no-U-turn and volume-preservation constraints. During warmup, the running average of $\alpha$ is recorded and used by dual averaging to adapt the step size toward a target acceptance probability.

The pre-training resulted parameter vector, denoted $\boldsymbol{\theta}^*$, is used both to initialize NUTS and to center the parameter prior. Accordingly, a Gaussian prior centered at the pre-trained weights is adopted:

$$p(\boldsymbol{\theta}) = N(\boldsymbol{\theta}^*, \sigma_p^2 \boldsymbol{I}) \qquad (16)$$

where $\sigma_p^2$ sets the prior variance and reflects the initial uncertainty around the pre-trained solution. The dataset $D$, used for Bayesian inference, consists of three types of residuals introduced in **Section 2.2**: the current conservation residuals $f^{(n)}$, the energy conservation residual $g^{(n)}$ and the measurement residuals $h^{(n,m)}$, each evaluated at their respective spatial points. Assuming these residuals are corrupted by independent Gaussian noise with standard deviations $\sigma_f, \sigma_g, \sigma_h$, chosen a priori and held fixed during inference. Hence, the total likelihood is defined as:

$$p(D|\boldsymbol{\theta}) = p(D_f|\boldsymbol{\theta}) \cdot p(D_g|\boldsymbol{\theta}) \cdot p(D_h|\boldsymbol{\theta}) \qquad (17a)$$

with each term given by:

$$p(D_f|\boldsymbol{\theta}) = \prod_{n=1}^{4} \prod_{i=1}^{N} \frac{1}{\sqrt{2\pi}\sigma_f} \exp\left(-\frac{f^{(n)}(x^{(i)})^2}{2\sigma_f^2}\right) \qquad (17b)$$

$$p(D_g|\boldsymbol{\theta}) = \prod_{n=1}^{4} \prod_{i=1}^{N} \frac{1}{\sqrt{2\pi}\sigma_g} \exp\left(-\frac{g^{(n)}(x^{(i)})^2}{2\sigma_g^2}\right) \tag{17c}$$

$$p(D_h|\boldsymbol{\theta}) = \prod_{n=1}^{4} \prod_{m=1}^{3} \frac{1}{\sqrt{2\pi}\sigma_h} \exp\left(-\frac{h^{(n,m)2}}{2\sigma_h^2}\right) \tag{17d}$$

These residuals enforce both the physical governing equations and agreement with measured data, and serve as the basis for probabilistic inference.

The Bayesian PINN comprises three independent MLPs that infer $T(x)$, $V(x)$ and $k(T)$, respectively. Hard constraints are used to enforce boundary conditions and the positivity of $k(T)$, and nondimensionalization is applied to enhance numerical stability. The posterior samples obtained via NUTS are used to construct credible intervals for the inferred thermal conductivity, enabling uncertainty-aware estimation that accounts for measurement noise.

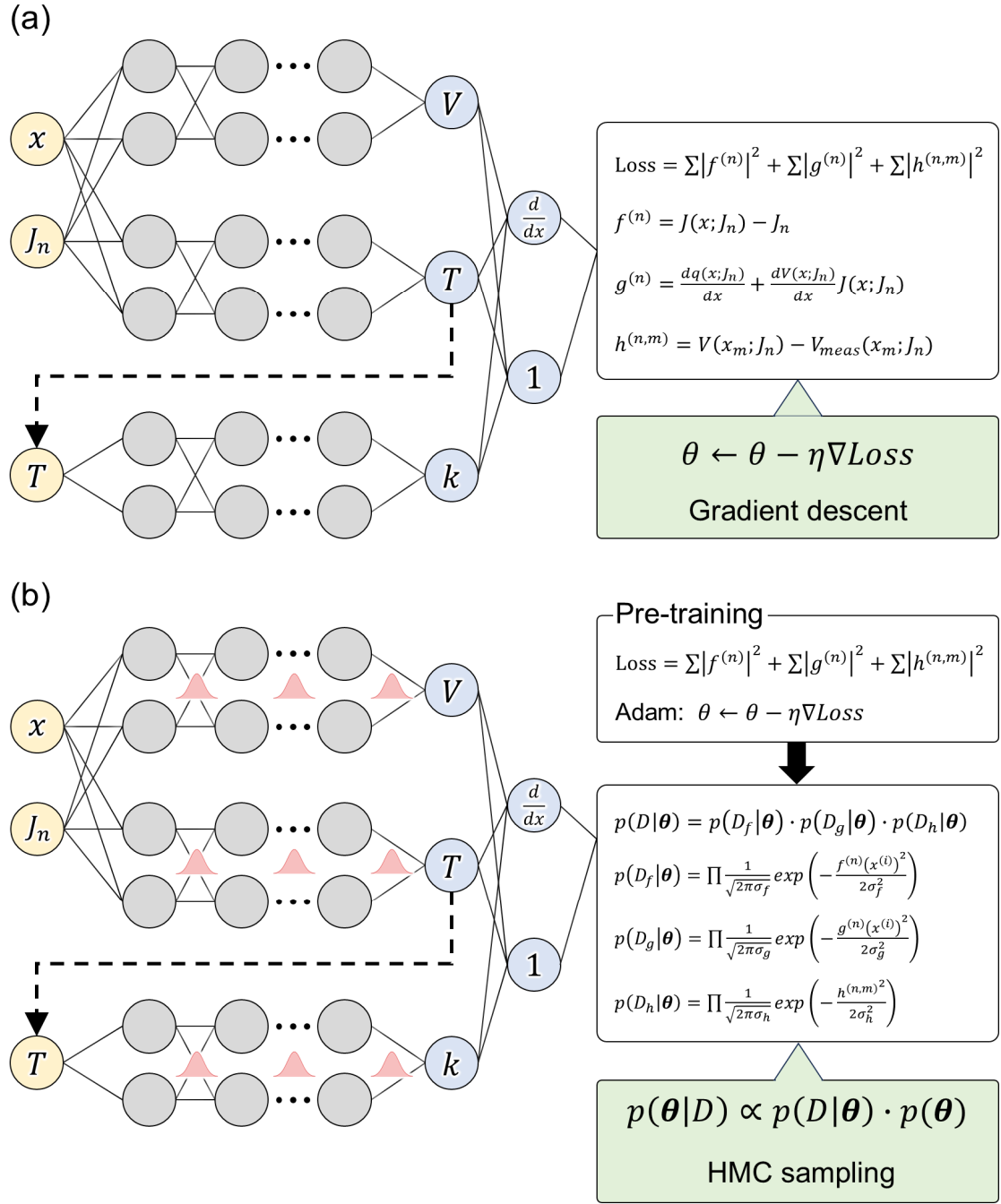

**Fig. 2. Architectures of the PINN and Bayesian PINN frameworks.** (a) PINN architecture for deterministic inference with noise-free data; the composite loss and a gradient-descent optimizer are shown., (b) Bayesian PINN architecture for probabilistic inference with noisy data; after pre-training using a gradient-descent optimizer, HMC sampling is run from the resulting initial state, and the likelihood terms used in sampling are illustrated.

## 4. Results and discussions

We evaluate the performance of the proposed PINN and Bayesian PINN frameworks on a p-type thermoelectric material. The material properties used in the simulation are summarized in **Fig. 3**. Ground-truth profiles for the temperature $T(x)$ and electric potential $V(x)$ were generated using high-fidelity numerical simulations performed in COMSOL Multiphysics. All physical quantities were nondimensionalized prior to training using the reference values $L_{ref} = 100 \, mm$, $T_{ref} = 350 \, K$, $k_{ref} = 1 \, W/m \cdot K$ and $\sigma_{ref} = 10,000 \, S/m$.

Both the PINN and Bayesian PINN employ the same network architecture, comprising three MLPs that infer $T(x)$, $V(x)$ and $k(T)$, respectively. Each MLP consists of 2 hidden layers with 10 neurons per layer. The PINN is trained using the Adam optimizer with a learning rate of $10^{-3}$, and the loss weight parameter $\lambda$ is set to 0.1 unless otherwise specified.

In case of Bayesian PINN, the network parameters are optimized with Adam optimizer for 20,000 epochs using a learning rate of 0.001. In Bayesian formulation, all network parameters are modeled as independent Gaussian variables with zero mean and unit standard deviation, i.e., $\sigma_p = 1$. Using NUTS, a total of 100 samples are drawn after a burn-in phase of 10 steps, with the target acceptance probability set to 0.8. The likelihood functions are constructed assuming independent Gaussian noise, with standard deviations $\sigma_f = 0.05$ and $\sigma_g = 0.5$ for the physics residuals, and $\sigma_h = 0.1$ mV for the measurement data. Here, $\sigma_f$ and $\sigma_g$ are defined in nondimensional form, consistent with the nondimensionalized formulation of the governing equations. In contrast, $\sigma_h$ is specified in its original dimensional form (prior to nondimensionalization), since the noise level was introduced directly into the measurement data before normalization. In practical settings, measurement instruments specify allowable noise levels. To incorporate this information into the framework, the likelihood standard deviations are set inversely proportional to the known noise level, so that higher noise increases the relative weight of the prior in the posterior. Thereby, the robustness of the framework to the measurement noise is enhanced.

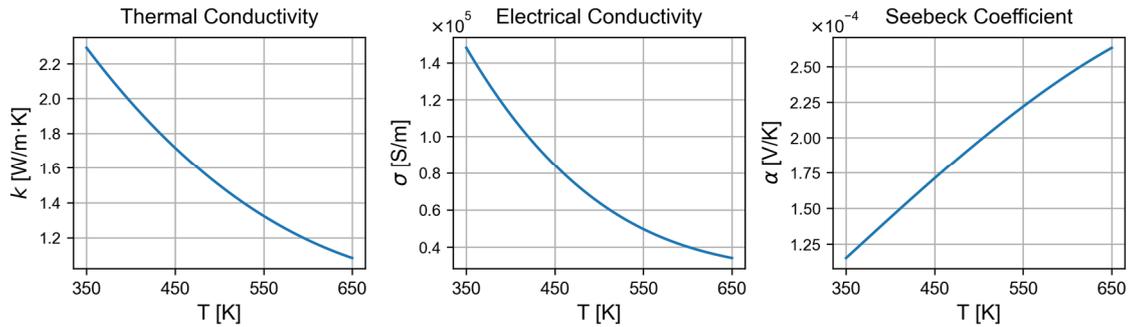

**Fig. 3. Ground-truth temperature-dependent material properties: thermal conductivity $k(T)$, electrical conductivity $\sigma(T)$, and Seebeck coefficient $\alpha(T)$**

**4.1) PINN results**

**4.1.1) Case with noise-free data**

**Fig. 4** presents the results of the inverse problem solved using the PINN framework under noise-free conditions. As shown in **Fig. 4(a)**, the physics loss and data loss both decrease and converge during training, indicating that the network successfully satisfies the governing equations while fitting the measurement data.

**Fig. 4(b)** shows the predicted temperature-dependent thermal conductivity $k(T)$, which closely matches the ground-truth over the entire temperature range. **Fig. 4(c)** compares the predicted and true profiles of temperature $T(x)$ and electric potential $V(x)$ for four different current loading cases ($J_n$=5, 10, 25, 50 $\mu A/mm^2$). In all cases, the predicted fields exhibit excellent agreement with the ground-truth, and the predicted electric potentials pass precisely through the measurement points.

These results demonstrate that the PINN framework can accurately solve the inverse problem using only sparse electric potential measurements obtained under multiple loading conditions. Remarkably, the temperature-dependent thermal conductivity $k(T)$ is successfully inferred without requiring any direct temperature measurements, underscoring the practicality and effectiveness of the PINN approach for experimental scenarios where temperature sensing is difficult or infeasible.

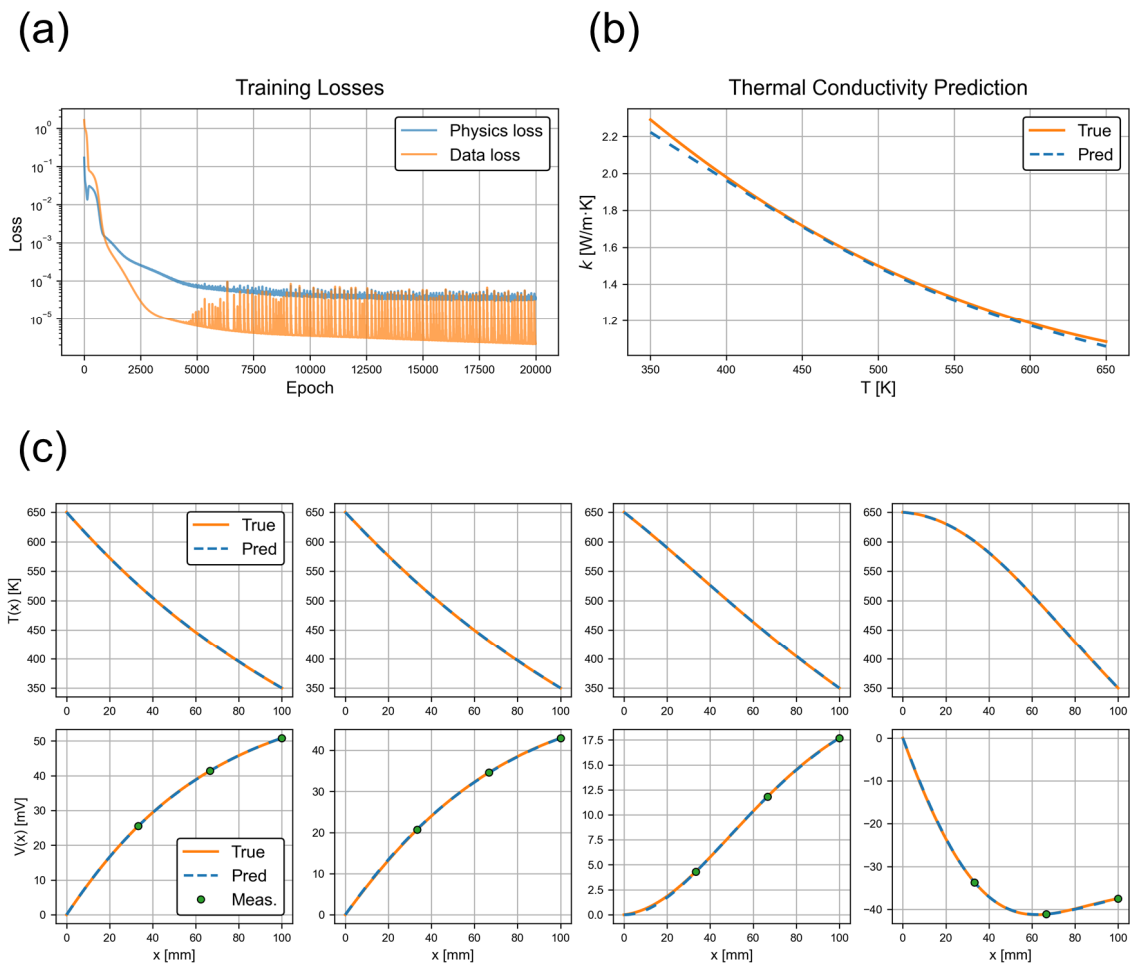

**Fig. 4. PINN results under noise-free conditions.** (a) Training losses with the two

components shown separately; physics loss and data loss, (b) Inferred thermal conductivity $k(T)$ compared with the ground-truth, (c) Predicted $T(x)$ and $V(x)$ profiles with their ground-truth curves for four current loading cases.

**4.1.2) Case with noisy data**

Since the proposed PINN framework infers the underlying physical fields solely from sparse measurement data, its performance is inherently sensitive to the quality of the input. To evaluate this sensitivity, artificial Gaussian noise was added to the electric potential measurements, following the formulation described in Eq. (5), and the impact on the inverse solution was assessed.

**Fig. 5** shows the performance of the PINN when Gaussian noise $\varepsilon$ with a standard deviation of 3 mV is added to the measurements, i.e., $\varepsilon \sim N(0, 3^2)$. As shown in **Fig. 5(a)**, the training loss converges sufficiently; however, the predicted thermal conductivity $k(T)$, presented in **Fig. 5(b)**, deviates significantly from the ground-truth. Furthermore, **Fig. 5(c)** illustrates that for all four current density loading cases, the predicted fields $T(x)$ and $V(x)$ are distorted to accommodate the noisy measurements, resulting in a significant degradation in the physical consistency of the predicted fields. These results indicate that the deterministic PINN tends to overfit noisy data in the absence of uncertainty modeling.

To further assess the impact of noise magnitude, we evaluated the model on noise-free data and on noisy observations with measurement-noise standard deviations of 1 mV and 5 mV, and compared the corresponding predictions, as shown in **Fig. 6**. In the noise-free case, the PINN accurately recovered the true physical fields. However, as the noise level increases, the predictive accuracy of $k(T)$ progressively deteriorated, and the estimated temperature profiles exhibited physically unrealistic behavior. For noisy datasets, PINNs can enhance predictive accuracy by adjusting the balance between the data and physics losses so that greater weight is assigned to the physics term; details are provided in **Supplementary A**.

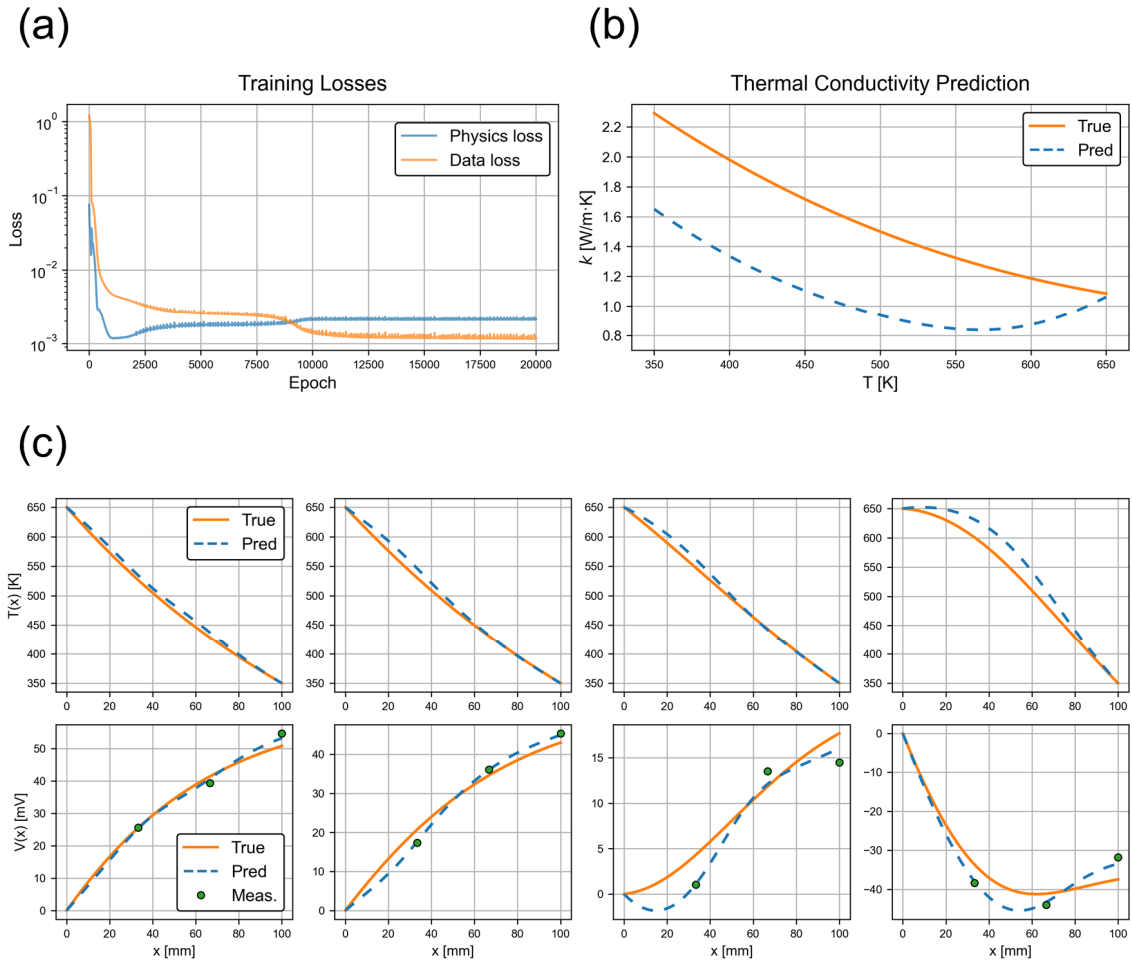

**Fig. 5. PINN results under $\varepsilon \sim N(0, 3^2)$.** (a) Training losses with the two components shown separately; physics loss and data loss, (b) Inferred thermal conductivity $k(T)$ compared with the ground-truth, (c) Predicted $T(x)$ and $V(x)$ profiles with their ground-truth curves for four current loading cases.

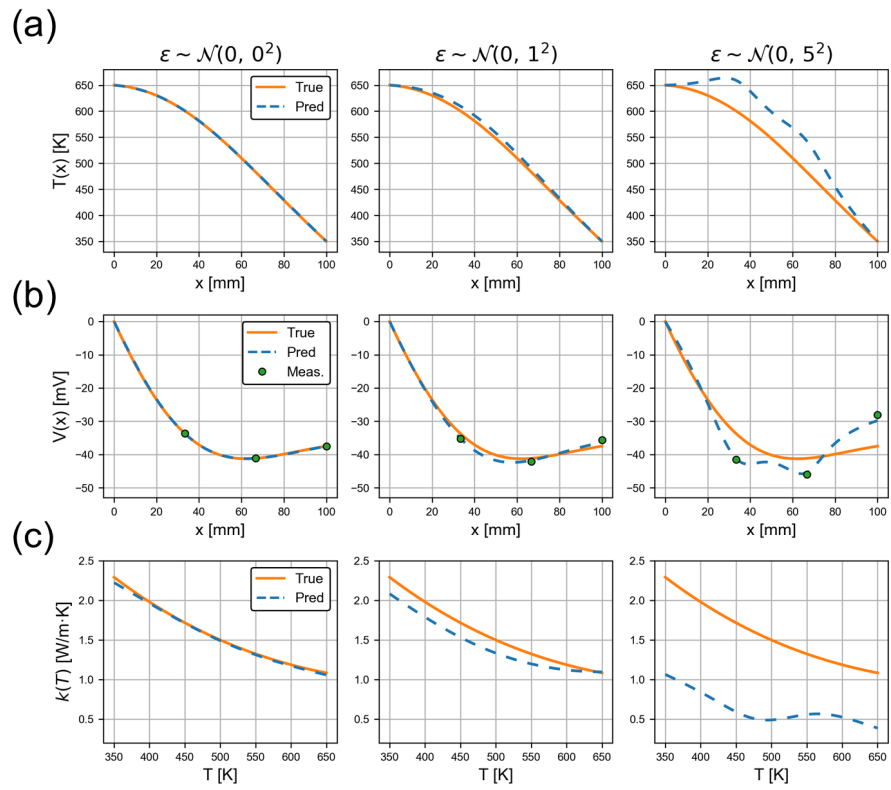

Fig. 6. PINN predictions at a fixed current density $J_n = 50m\ A/m^2$ under varying observation noise levels. Predicted values are shown alongside the ground-truth curves. (a) Temperature $T(x)$, (b) Electric potential $V(x)$; green circles indicate the measured voltages with assigned Gaussian noise, (c) Inferred thermal conductivity $k(T)$.

### 4.2) Bayesian PINN results

To overcome the limitations of the PINN in the presence of measurement noise, we adopt the Bayesian PINN framework, which enables uncertainty-aware inference by modeling neural network parameters as probability distributions. This approach provides estimates of $k(T)$ along with a quantification of the uncertainty associated with noisy observations. The credible intervals obtained from the posterior distribution offer a more interpretable assessment of the reliability of the inferred solution.

**Fig. 7** shows the posterior mean and 95% credible intervals of $k(T)$ under different levels of Gaussian noise ($\sigma = 1,3,5$ mV). As the noise increases, the mean prediction generally remains close to the ground-truth. However, across all three cases, noticeable deviations appear in the lower temperature region even when the overall error is small at $\sigma = 1$mV. This bias likely arises because the cold-end region (near 350 K) is farther from both the fixed Dirichlet boundary at the hot end (T = 650K, where V = 0) and from the measurement point, resulting in weaker constraints and higher local uncertainty in the inference. Additionally, as the noise level increases, the error relative to the ground-truth grows, and the credible interval width, which reflects uncertainty across the temperature range, increases as well. To make this trend more intuitive, **Fig. 8** reports, for each noise level, the mean absolute error and the mean 95% credible-interval width computed over the temperature range. Consequently, both the absolute error and the width of the credible intervals obtained from the Bayesian PINN increase with the noise level, exhibiting a consistent common upward trend as input noise rises.

To further evaluate the prediction performance, **Fig. 9** illustrates the inferred temperature $T(x)$, electric potential $V(x)$, and the nonlinear deviation $T(x) - T_{linear}(x)$ at a fixed current density of $J_n = 50 \mu A/mm^2$, under varying noise magnitudes. Here, $T_{linear}$ denotes the baseline temperature profile obtained by solving the steady-state heat equation assuming constant thermal conductivity, i.e., $T_{linear}(x) = T_h - \frac{T_h - T_c}{L}x$. This reference solution serves to isolate the nonlinearity caused by the temperature dependence of $k(T)$ and thermoelectric effects.

As shown in **Fig. 9**, the predictive accuracy of $T(x)$ remains high regardless of the noise level. The credible intervals are uniformly narrow over the domain, indicating well-constrained and confident predictions. For $V(x)$, predictive accuracy remains high across noise levels. As the noise level increases, the posterior variance widens near the measurement points, reflecting the higher observational noise and providing an uncertainty-aware indicator of potential error. Regarding the nonlinear aspects of $T(x)$, the credible intervals widen with increasing noise, similar to $V(x)$. However, prediction accuracy is slightly lower, likely because the pronounced nonlinearity reduces identifiability.

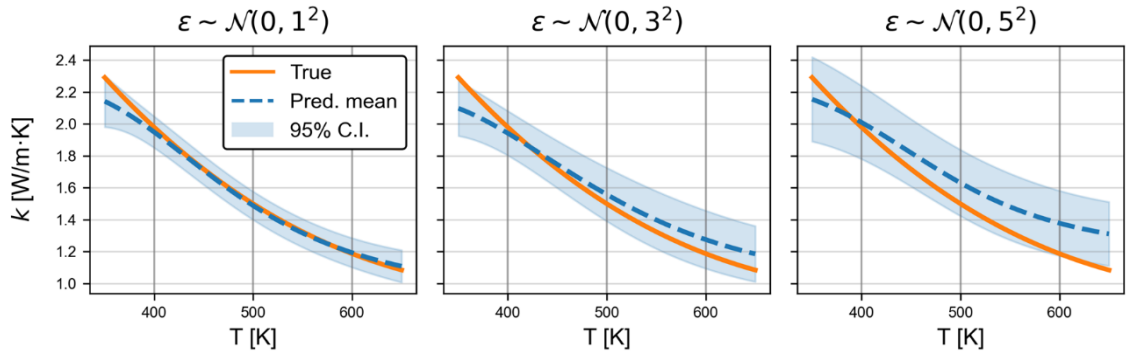

**Fig. 7. Predicted mean and 95% credible intervals of $k(T)$ inferred by Bayesian PINN under different Gaussian noise levels.**

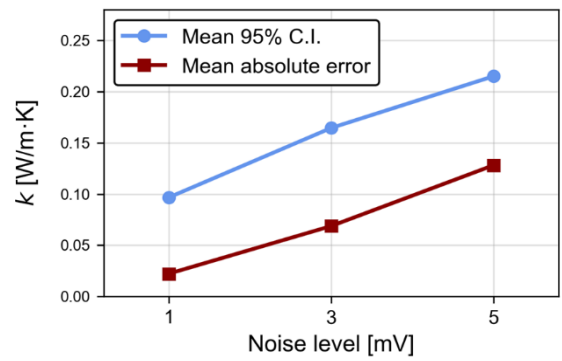

**Fig. 8.** Pointwise comparison between the absolute error of the Bayesian PINN posterior mean and ground-truth of $k(T)$ and the width of the 95% credible intervals under varying Gaussian noise.

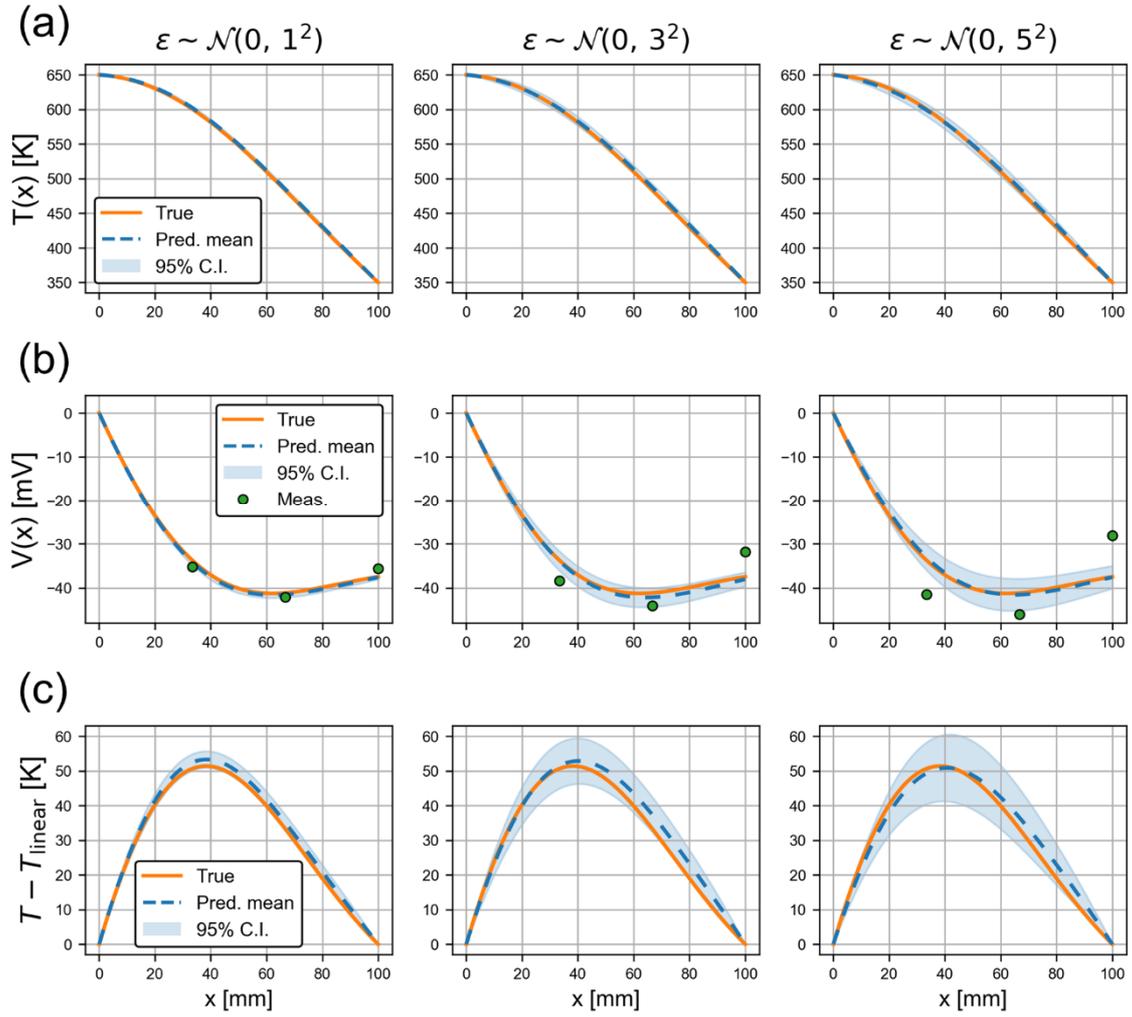

**Fig. 9. Bayesian PINN predictions at a fixed current density $J_n = 50m\ A/m\ m^2$ under varying observation noise levels. The predicted mean with its 95% credible intervals are shown alongside the ground-truth curves.** (a) Temperature $T(x)$, (b) Electric potential $V(x)$; green circles indicate the measured voltages with assigned Gaussian noise, (c) Nonlinear temperature deviation $T(x) - T_{linear}(x)$

### 4.3) Comparison between Bayesian PINN and PINN for noisy data

While deterministic PINN can recover accurate physical fields under ideal, noise-free conditions, their performance rapidly deteriorates in the presence of measurement noise.

Across all tested loading scenarios, the PINN exhibited significant sensitivity to corrupted voltage data, resulting in distorted predictions of $T(x)$ and $V(x)$, and unphysical behavior in the inferred $k(T)$. These limitations stem from the lack of a principled mechanism to model or quantify uncertainty, which causes the deterministic network to overfit noisy inputs and undermines its reliability.

In contrast, the Bayesian PINN demonstrates enhanced robustness, interpretability, and diagnostic capability. By treating network parameters as probabilistic variables, the Bayesian PINN generates posterior distributions over the inferred quantities, allowing for a natural quantification of uncertainty. As shown in **Fig. 7** and **9**, the Bayesian PINN remains robust even under high-noise conditions. Moreover, the credible intervals become wider as the noise level and error increase, allowing the model to not only provide predictions but also offer insight into the magnitude of input uncertainty and the associated risk of predictive failure, purely from inference. As shown in **Fig. 7** and **8**, although the overall trend is consistent, with both the absolute error and the width of the credible intervals increasing with the noise level, the posterior uncertainty does not always capture local absolute errors within individual cases. This indicates a limitation in local calibration. This behavior arises from the characteristics of the problem setting: the simplified 1D approximation limits the influence of noise, and the inverse problem is governed by strongly coupled equations with well-constrained boundary conditions. As a result, the effect of noise is diluted and weakened as it propagates from $V_{meas}$ to $k(T)$.

In summary, the Bayesian PINN offers significant advantages over its deterministic counterpart. It improves robustness to measurement noise and delivers meaningful uncertainty quantification, particularly within the proposed problem setting. These features make the

Bayesian PINN a more powerful and reliable framework for physics-informed inference in uncertain, realistic, and practical environments.

## 5. Conclusion

This study presents a Bayesian PINN-based framework for estimating the temperature-dependent thermal conductivity $k(T)$ of thermoelectric materials without requiring direct temperature measurements. The approach leverages the coupled thermoelectric governing equations to formulate an inverse problem, where $k(T)$ is inferred from sparse electric potential measurements obtained under multiple current loading conditions.

In the absence of noise, conventional PINN successfully infers $k(T)$, demonstrating the feasibility of reconstructing spatially varying thermal properties using only electric potential data. However, the results also indicate that the approach is sensitive to both measurement noise and the selection of the loss weighting parameter $\lambda$, which may affect the stability and reliability of the solution in practical applications.

To address these limitations, the framework is extended to a Bayesian PINN, wherein the neural network parameters are modeled as probability distributions. This formulation enables posterior inference of $k(T)$ via HMC sampling, yielding not only representative estimates but also credible intervals that quantify uncertainty arising from noisy and limited data. Under realistic noise levels, the Bayesian PINN provides reliable predictions, and the associated credible intervals reflect the input noise level and enable risk-aware interpretation by indicating where uncertainty is high and prediction failures are more likely.

**CRediT authorship contribution statement**

**Hyeonbin Moon:** Conceptualization, Methodology, Software, Formal analysis, Validation, Visualization, Writing – original draft, Writing – review & editing. **Hanbin Cho:** Methodology, Software, Investigation, Data curation, Visualization, Writing – review & editing. **Wabi Demeke:** Validation, Data curation, Writing – review & editing. **Byungki Ryu:** Conceptualization, Methodology, Supervision, Writing – review & editing. **Seunghwa Ryu:** Conceptualization, Supervision, Project administration, Funding acquisition, Writing – review & editing.

**Declaration of Competing Interest**

The authors declare that they have no known competing financial interests or personal relationships that could have appeared to influence the work reported in this paper

**Data availability**

Data will be made available on request.

**Acknowledgements**

This work was supported by the InnoCORE program (N10250154) and the National Research Foundation of Korea (RS-2023-00247245) grant of the Ministry of Science and ICT.

**Supplementary Information**

# Thermal Conductivity Estimation of Thermoelectric Materials with Uncertainty Quantification Using Bayesian Physics-Informed Neural Networks

Hyeonbin Moon[1†], Hanbin Cho [1†], Wabi Demeke[1], Byungki Ryu[2*] and Seunghwa Ryu[1,3*]

**Affiliations**

[1] Department of Mechanical Engineering, Korea Advanced Institute of Science and Technology (KAIST), 291 Daehak-ro, Yuseong-gu, Daejeon 34141, Republic of Korea

[2] Energy Conversion Research Center, Korea Electrotechnology Research Institute (KERI), 12 Jeongiui-gil, Seongsan-gu, Changwon-si, Gyoengsangnam-do 51543, Republic of Korea

[3] KAIST InnoCORE PRISM-AI Center, Korea Advanced Institute of Science and Technology (KAIST), Daejeon, 34141, Republic of Korea

[†]These authors contributed equally to this work

[*]Corresponding author: ryush@kaist.ac.kr (Seunghwa Ryu), byungkiryu@keri.re.kr (Byungki Ryu)

**Supplementary A. PINN prediction results of $k(T)$ under varying loss weight**

While adjusting the physics-data trade off via the loss weight $\lambda$ can improve the robustness of PINNs, the effect is highly problem-dependent and does not generalize reliably. In our setting, we studied the sensitivity of the inferred $k(T)$ to $\lambda$ by sweeping it over orders of magnitude. The loss function was defined as

$$\mathcal{L}(\boldsymbol{\theta}; \lambda) = \mathcal{L}_{data}(\boldsymbol{\theta}) + \lambda \mathcal{L}_{phys}(\boldsymbol{\theta}) \qquad (A.1)$$

where $\mathcal{L}$ refers the total loss and $\theta$ refers the parameter vector. $\mathcal{L}_{data}$ measures the data misfit between predicted and measured voltages and $\mathcal{L}_{phys}$ penalizes violations of the governing equations. $\lambda$ works as weight between the two loss terms. As shown in **Fig. A.1**, when $\lambda$ is small (e.g., 0.1 to 1), the model places excessive emphasis on fitting the noisy measurements, leading to poor recovery of the true thermal conductivity $k(T)$. As $\lambda$ increases, the model prioritizes the enforcement of physical laws, and the prediction improves, with the best match to the ground truth occurring at $\lambda=10$. However, further increasing $\lambda$ beyond this point once again degrades performance, as the model begins to underfit the available data. These results suggest that while careful tuning of $\lambda$ can partially mitigate noise sensitivity, PINN lack a principled mechanism for quantifying how uncertainty in the data propagates to the inferred material properties, thereby motivating a Bayesian approach.

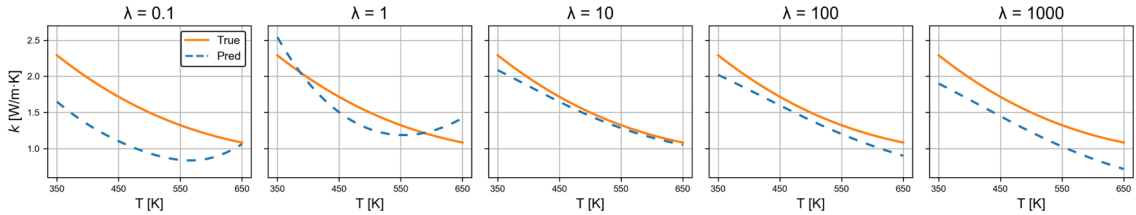

**Fig. A.1. Effect of the physics loss weight $\lambda$ on the inferred $k(T)$. Curves show the

learned $k(T)$ for representative $\lambda$ values; the ground truth $k(T)$ is plotted for reference.